%% file: draft.tex
\documentclass[twocolumn,prd,noshowpacs,nofootinbib,amsmath,amssymb,superscriptaddress]{revtex4}
\usepackage{graphicx,epsfig,psfrag,bm,amssymb}
\usepackage{dcolumn}
\usepackage{bm}
\usepackage{mciteplus}
\usepackage{color}
\usepackage{mathrsfs,amsfonts,hepunits, color}
\usepackage{mciteplus}
\usepackage{tikz}
\usepackage{slashed}
\usepackage{cancel}
\usetikzlibrary{arrows,shapes}
\usetikzlibrary{trees}
\usetikzlibrary{matrix,arrows} 				
\usetikzlibrary{positioning}				
\usetikzlibrary{calc,through}				
\usetikzlibrary{decorations.pathreplacing}  
\usepackage{pgffor}							

\usetikzlibrary{decorations.pathmorphing}	
\usetikzlibrary{decorations.markings}
\usetikzlibrary{snakes}

\tikzset{
    vector/.style={decorate, decoration={snake}, draw},
	provector/.style={decorate, decoration={snake,amplitude=2.5pt}, draw},
	antivector/.style={decorate, decoration={snake,amplitude=-2.5pt}, draw},
    fermion/.style={draw, postaction={decorate},
        decoration={markings,mark=at position .55 with {\arrow[draw]{>}}}},
    fermionbar/.style={draw, postaction={decorate},
        decoration={markings,mark=at position .55 with {\arrow[draw=black]{<}}}},
    fermionnoarrow/.style={draw},
    gluon/.style={decorate, draw,decoration={coil,amplitude=4pt, segment length=6pt}, line width=1},
    scalar/.style={dashed,draw, postaction={decorate},
        decoration={markings,mark=at position .55 with {\arrow[draw]{>}}}},
    scalarbar/.style={dashed,draw, postaction={decorate},
        decoration={markings,mark=at position .55 with {\arrow[draw]{<}}}},
    scalarnoarrow/.style={dash pattern = on 6 pt off 3 pt,draw},
    electron/.style={draw, postaction={decorate},
        decoration={markings,mark=at position .55 with {\arrow[draw]{>}}}},
	bigvector/.style={decorate, decoration={snake,amplitude=4pt}, draw},
	vectorscalar/.style={loosely dotted,draw, postaction={decorate}},
}
\RequirePackage{xspace}
\usepackage{relsize}
\usepackage{slashed}

\newcommand{\be}{\begin{eqnarray}}
\newcommand{\ee}{\end{eqnarray}}
\def\lsim{\mathrel{\rlap{\lower4pt\hbox{\hskip 0.5 pt$\sim$}}
    \raise1pt\hbox{$<$}}}                
\def\gsim{\mathrel{\rlap{\lower4pt\hbox{\hskip1pt$\sim$}}
    \raise1pt\hbox{$>$}}} 
    
\def\lsim{\mathrel{\rlap{\lower4pt\hbox{\hskip1pt$\sim$}}
    \raise1pt\hbox{$<$}}}
\def\gsim{\mathrel{\rlap{\lower4pt\hbox{\hskip1pt$\sim$}}
    \raise1pt\hbox{$>$}}}

\newcommand{\bi}{\begin{itemize}}
\newcommand{\ei}{\end{itemize}}
\newcommand{\bea}{\begin{eqnarray}}
\newcommand{\eea}{\end{eqnarray}}

\newcommand{\benum}{\begin{enumerate}}
\newcommand{\eenum}{\end{enumerate}}

\begin{document}

\title{
Multilepton and Lepton Jet Probes of Sub-Weak-Scale Right-Handed Neutrinos}

\author{Eder Izaguirre}
\author{Brian Shuve}
 \affiliation{Perimeter Institute for Theoretical Physics, Waterloo, Ontario, Canada    }

\begin{abstract}
We propose new searches that exploit the unique signatures of decaying sterile neutrinos with masses below $M_W$ at the LHC, where they can be produced in rare decays of Standard Model gauge bosons. We show that for few-\GeV-scale sterile neutrinos, the LHC experiments can probe mixing angles at the level of $10^{-4}-10^{-3}$ through powerful searches that look for a prompt lepton in association with a displaced lepton jet. For higher-mass sterile neutrinos, {\it i.e.} $M_N \gsim 15\,\, \GeV$, Run II can explore similarly small mixing angles in prompt multilepton final states. This represents an improvement of up to two orders of magnitude in sensitivity to the sterile neutrino production rate.
\end{abstract}

\maketitle

%
%
\section{Introduction}
\label{sec:intro}

\input{introduction}

\section{Simplified Model}\label{sec:simplified_model}

\input{simplified_model}

\section{Lepton Jets from RH Neutrinos}\label{sec:lepton_jets}

\input{lepton_jet}

\section{Prompt Trilepton Searches for RH Neutrinos}\label{sec:prompt}

\input{prompt}

\section{Discussion and Conclusions}\label{sec:concl}

\input{concl}

\appendix

\section{Fake Lepton Simulation}\label{app:fakesim}
\input{fakesim}

\bibliography{draft}

\end{document}

%% file: introduction.tex
Run I of the Large Hadron Collider (LHC) proved a success by any measure, leading to the discovery of the Standard Model (SM) Higgs boson \cite{Aad:2012tfa,Chatrchyan:2012ufa} after decades of pursuit by previous experiments. Moreover, the successes of the SM have now been extended by the LHC experiments to energy scales well above a TeV, as evidenced by the exquisite agreement between data and theory. However, there are key outstanding missing pieces in the SM, chief among them the origin of neutrino masses \cite{Agashe:2014kda}, the identify of dark matter, and the dynamics responsible for the baryon asymmetry \cite{Planck:2015xua}. Remarkably, a minimal extension of the SM with three gauge-singlet, ``sterile'' right-handed (RH) neutrinos can resolve all of these problems if the three right-handed neutrinos all lie below the weak scale. Known as the neutrino minimal SM ($\nu$MSM) \cite{Asaka:2005an,Asaka:2005pn,Canetti:2012kh}, the new sterile states lie within kinematic reach of various laboratory probes \cite{Gorbunov:2007ak}, leading to the exciting possibility that existing and upcoming experiments could shed light on the particles responsible for neutrino masses  \cite{Minkowski:1977sc,Yanagida:1979as,Mohapatra:1979ia,GellMann:1980vs,Schechter:1980gr}, dark matter \cite{Dodelson:1993je,Shi:1998km,Abazajian:2001nj,Asaka:2005an,Laine:2008pg}, and baryogenesis \cite{Akhmedov:1998qx,Asaka:2005pn,Shaposhnikov:2008pf,Asaka:2010kk,Canetti:2010aw,Asaka:2011wq,Canetti:2012kh,Drewes:2012ma,Shuve:2014zua} \footnote{A recent review of the collider phenomenology of RH neutrinos can be found in Ref.~\cite{Deppisch:2015qwa}.}.

\begin{figure}[t]
\centering
\includegraphics[width=0.2 \textwidth ]{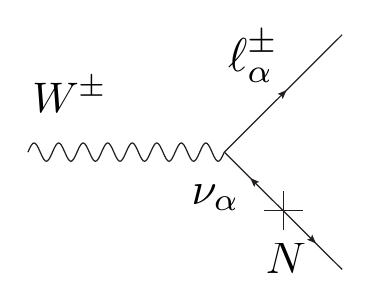}\hspace{0.5cm}
\includegraphics[width=0.2 \textwidth ]{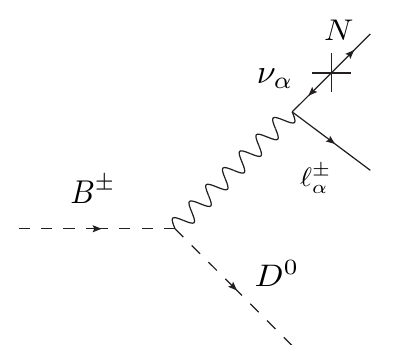}
\caption{Feynman diagrams showing possible production mechanisms of a right-handed neutrino, $N$, via mixing with SM neutrinos in (left) the decay of a SM gauge boson ($W^\pm \rightarrow \ell^\pm N$); (right) the decay of a $B$ meson.}
\label{fig:N_prod}
\end{figure}

After electroweak symmetry breaking, the $\nu$MSM parameters consist of the masses of the RH neutrinos, $N_I$, and a mixing angle between each flavor  $I$ of sterile neutrino and each flavor $\alpha$ of active neutrino, $V_{\alpha N_I}$. The interactions between $N_I$ and SM fields are then completely determined by the mixing between neutrinos; $N_I$ can be produced through any interaction involving SM neutrinos, with two examples shown in Fig.~\ref{fig:N_prod}. The matrix element is the same as for the corresponding process for SM neutrinos, with an extra multiplicative factor of $V_{\alpha N_I}$. The same interactions also allow $N$ to decay to SM states.  As we elaborate in Section \ref{sec:simplified_model}, we focus on the mixing angle with the muon neutrino, $V_{\mu N_I}$.

\begin{figure}[t]
\centering
\includegraphics[width=0.5 \textwidth ]{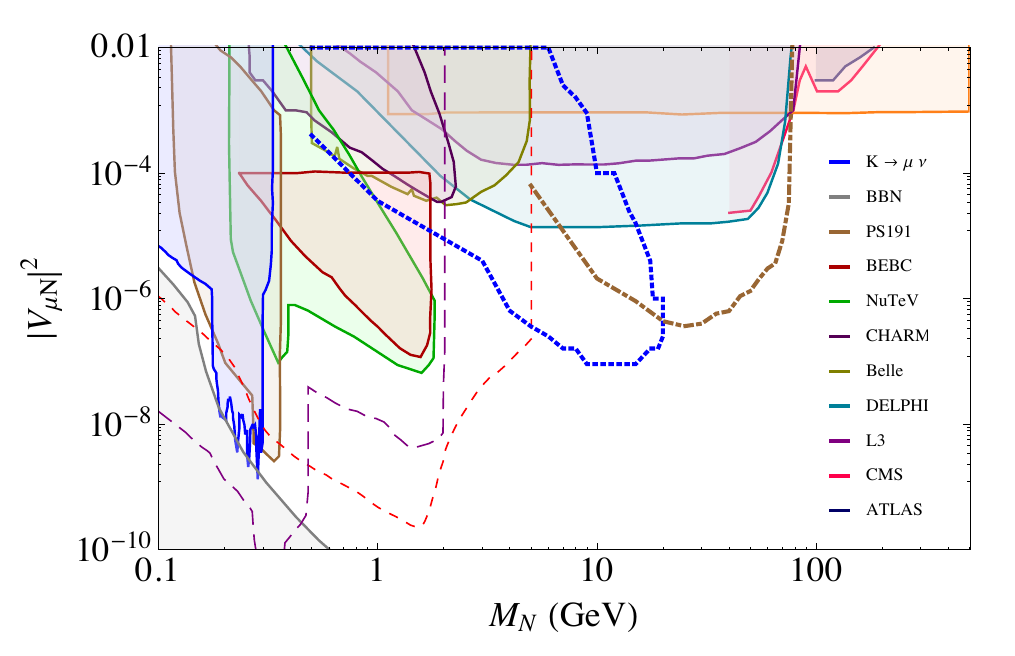}
\caption{Constraints (shaded regions bounded by solid lines) on the right-handed neutrino mass, $M_N$, and squared mixing angle, $|V_{\mu N}|^2$ \cite{Deppisch:2015qwa}. We show the estimated $2\sigma$ reach of our two proposed searches with $\sqrt{s}=13$ TeV and $300\,\,\mathrm{fb}^{-1}$:~the displaced lepton jet search discussed in Section \ref{sec:lepton_jets} (blue, dotted), and the prompt trilepton search discussed in Section \ref{sec:prompt} (brown, dot-dashed).  The simplified model is defined in Section \ref{sec:simplified_model}. For comparison, the reach of the proposed DUNE/LBNF (purple, long dashed) and SHiP (red, short dashed) experiments are also shown.}
\label{fig:constraint_plot}
\end{figure}

The current constraints for a single RH neutrino mixing with $\nu_\mu$ are shown in Fig.~\ref{fig:constraint_plot} \cite{Yamazaki:1984sj,Bernardi:1985ny,Grassler:1986vr,Adriani:1992pq,Vilain:1994vg,Abreu:1996pa,Vaitaitis:1999wq,Gorbunov:2007ak,Boyarsky:2009ix,Ruchayskiy:2012si,ATLAS-CONF-2012-139,Liventsev:2013zz,Artamonov:2014urb,Khachatryan:2015gha,Deppisch:2015qwa}, with the least explored parameter space being $M_N > m_b$. The proposed SHiP \cite{Bonivento:2013jag} and DUNE/LBNF experiments \cite{Adams:2013qkq} are expected to greatly improve sensitivity for $N$ below the charm mass (and, to a lesser extent, the $b$ mass), as shown in Fig.~\ref{fig:constraint_plot}. Above the $b$ mass, the dominant production mechanism for $N$ is via the decay of gauge bosons, $W^\pm \rightarrow \ell^\pm N$ and $Z\rightarrow \nu N$. Experiments such as SHiP lack the center-of-mass energy needed to produce on-shell gauge bosons, thus experiments at high-energy colliders such as LEP \cite{Adriani:1992pq,Abreu:1996pa} and the LHC \cite{ATLAS-CONF-2012-139,Khachatryan:2015gha} are needed to discover the RH neutrinos. LEP was sensitive to $N$ production via the process $Z\rightarrow \nu N$, but its sensitivity was limited to $|V_{\alpha N_I}|^2\lsim 10^{-5}$ for $M_N\lesssim M_Z$. 

The LHC features large luminosity and cross sections for $N$ production via the decay of SM vector bosons, but the soft decay products of $W/Z$ could be mimicked by various reducible and irreducible SM processes. Furthermore, for $M_N\ll M_W$, the decay products of $N$ are collimated and may fail isolation requirements in conventional searches. Therefore, carefully targeted analyses are needed to extract the sterile neutrino signatures. We perform a comprehensive study of signatures of sterile neutrinos from $W^\pm $ decay at the LHC in the kinematic regime $M_N< M_W$. We propose new searches for two generic parts of parameter space, namely the case where $N$ decays at a displaced vertex, and the case where $N$ decays promptly. In the low-mass regime, $M_N\lesssim15$ GeV, the sterile neutrino is both long-lived and boosted; its leptonic decays will typically fail standard lepton isolation, but instead give a distinctive signature --- that of a displaced lepton jet  (see \cite{Strassler:2006im,Strassler:2006ri,Han:2007ae,Gopalakrishna:2008dv,ArkaniHamed:2008qp,Baumgart:2009tn,Cheung:2009su,Falkowski:2010cm} for a discussion of lepton jets in other contexts). In the higher-mass regime, 15 GeV $\lesssim M_N \lesssim M_W$, the sterile neutrino decays promptly and to three separately resolved objects. We find here that targeted trilepton searches can uncover sterile Majorana neutrinos in spite of significant SM backgrounds from lepton fakes. The trilepton analysis performance is more powerful than the only current LHC search for $M_N<M_W$, a $W^\pm\rightarrow \mu^\pm (N\rightarrow \mu^\pm jj)$ analysis from CMS \cite{Khachatryan:2015gha}. The estimated reach of our proposed searches at LHC13 with $300\,\,\mathrm{fb}^{-1}$ of integrated luminosity is shown in Fig.~\ref{fig:constraint_plot}.

This work differs from existing proposals for sterile neutrino searches at the LHC in displaced vertices \cite{Helo:2013esa} by exploiting the distinct, boosted kinematics of the final state for $M_N \ll M_W$. In particular, for low-mass displaced $N$, we advocate looking for the distinct signature of a prompt lepton in association with a displaced lepton jet.  For higher-mass $N$, when its decays are prompt, our proposals for prompt trilepton searches could cover new ground by improving sensitivity to $|V_{\mu N}|^2$ in the $m_b < M_N< M_W$ range by up to two orders of magnitude over existing bounds. This complements existing proposals for LHC searches in $\mu^\pm\mu^\pm jj$  and trilepton final states that typically target $M_N > M_Z$ \cite{Keung:1983uu,Pilaftsis:1991ug,Datta:1993nm,Almeida:2000pz,Panella:2001wq,Han:2006ip,Bray:2007ru,delAguila:2007em,delAguila:2008cj,delAguila:2008hw,Atre:2009rg,delAguila:2009bb,Perez:2009mu,Chen:2011hc,Das:2012ze,Das:2014jxa,Bambhaniya:2014kga} (although the reach of $\mu^\pm\mu^\pm jj$ searches at the Tevatron and LHC for $M_N<M_W$ are  examined in \cite{Han:2006ip,delAguila:2007em,Atre:2009rg}). In this high-mass scenario, $N$ are produced through off-shell gauge bosons and thus their decay products exhibit very different kinematics from $M_N < M_W$. Unlike many trilepton searches that are targeted at discovering Dirac neutrinos or distinguishing between various RH neutrino models \cite{delAguila:2008cj,delAguila:2008hw,delAguila:2009bb,Chen:2011hc,Das:2012ze,Das:2014jxa,Bambhaniya:2014kga,Dermisek:2014qca}, we exploit the lepton-number-violating character of the Majorana neutrinos in fully leptonic decays to improve sensitivity to $N$.  Parts of the parameter space that cannot be discovered at the LHC with our proposals can potentially be probed in future lepton colliders, such as the FCC-$ee$ \cite{Blondel:2014bra,Banerjee:2015gca}.

The rest of this article is structured as follows. We first describe in Section \ref{sec:simplified_model} the simplified model under consideration, along with the primary RH neutrino production and decay modes. In Section \ref{sec:lepton_jets}, we discuss the displaced lepton jet signature of RH neutrinos, and in Section \ref{sec:prompt}, the prospects for prompt multilepton searches. Finally, we conclude with the discussion in Section \ref{sec:concl}.

%% file: simplified_model.tex
The simplified model we consider is the SM with the addition of a single new RH neutrino $N$ state:
\be
\mathcal L \supset F_{\alpha}\bar L_{\alpha}H N + \frac{1}{2}M\,\bar N^{\rm c} N,
\ee
where $L_\alpha$ is the SM lepton doublet of flavor $\alpha$, $F_{\alpha}$ is a Yukawa coupling, and $H$ is the Higgs field. In more realistic models of neutrino masses, such as the $\nu$MSM, there are three flavors of $N$, and the above Lagrangian is supplemented with appropriate indices for the RH neutrinos. However, typically only one flavor of $N$ is produced at a time in any given interaction at an experiment, and so we can consider them independently in evaluating discovery potential. Similarly, we assume for simplicity that $N$ couples only to a single lepton doublet $\alpha$.

After electroweak symmetry breaking, the RH neutrino acquires a Dirac mass in addition to the Majorana mass and mixes with the SM neutrinos. In the limit $M \gg F_\alpha\langle H\rangle$, the mass eigenstates are
\bea
M_{\nu_\alpha}& \sim& \frac{F_\alpha^2\langle H\rangle^2}{M}\label{eq:seesaw}\\
M_N &\sim & M.
\eea
This is the conventional see-saw mechanism for neutrino masses.

In the same limit, the mixing angles between the active and sterile species have the scaling
\be
V_{\alpha N}\sim \frac{F_{\alpha }\langle H\rangle}{M}\ll 1.
\ee
Thus, any coupling $g_\nu$ involving the SM neutrino gauge eigenstate $\nu_\alpha$ will, after electroweak symmetry breaking, also lead to a coupling to $N$ equal to $g_\nu\,V_{\alpha N}$. For two examples, see Fig.~\ref{fig:N_prod}.

The see-saw relation, Eq.~(\ref{eq:seesaw}), gives a na\"ive prediction for $V_{\alpha N}$ as a function of $M_N$ for fixed $M_\nu$. However, $V_{\alpha N}$ depends linearly on $F_\alpha$, while $M_{\nu_\alpha}$ depends on $F_{\alpha}^2$, and so a cancellation between real and complex parts of $F_\alpha$ can give rise to a much larger angle than the na\"ive scaling suggests \cite{Casas:2001sr}\footnote{In Ref.~\cite{Drewes:2015iva}, it was found that this cancelation may not be radiatively stable in the minimal theory for $M_N\gg$ GeV, although theories with additional lepton number symmetries \cite{Shaposhnikov:2006nn} may avoid this problem.}. Additionally, extended theories such as the inverse see-saw mechanism \cite{Mohapatra:1986bd} can give rise to $V_{\alpha N}$ much larger than the na\"ive see-saw prediction; therefore, we treat $V_{\alpha N}$ as a free parameter in the simplified model.

In this paper, we consider the specific simplified model where $N$ mixes predominantly with the muon doublet, $L_\mu$. This is the simplified model constrained in current LHC searches for RH neutrinos \cite{ATLAS-CONF-2012-139,Khachatryan:2015gha}, and is most amenable for detection at colliders due to the clean, muon-rich signatures. While we also advocate for direct probes of $V_{eN}$ and $V_{\tau N}$ at the LHC, these suffer from larger mis-tag rates than muons\footnote{The sensitivity to $V_{eN}$ of current and upcoming   neutrinoless double-beta decay experiments \cite{Ackermann:2012xja} is also expected to be competitive with, if not better than, the reach of high-energy colliders such as the LHC.}, and due to the difficulties of estimating backgrounds, the optimization of these studies should be undertaken by the LHC experimental collaborations. We show the current constraints on the simplified model parameter space $(M_N,V_{\mu N})$ in Fig.~\ref{fig:constraint_plot}.

The weakest limits are for $M_N\gtrsim m_c$, and we therefore focus on this part of parameter space. We further narrow our  investigation to the promising $M_N < M_W$ range, due to the large production rate of $N$ from gauge boson and/or meson decays.\\

\noindent {\bf Production:} In our simplified model, $N$ production and decay proceed entirely through the mixing with the $\nu_\mu$. The most easily detected production modes for $N$ are:
\benum
\item In $W$ boson decays, $W^\pm\rightarrow N\mu^\pm$\footnote{We do not consider the sub-dominant process $Z\rightarrow \nu N$ due to the lower rate and the lack of a prompt lepton for passing the trigger. However, this process can become relevant at a future $e^+e^-$ collider \cite{Blondel:2014bra,Banerjee:2015gca}.};
\item When kinematically allowed, in heavy-flavor hadron decays such as $B^\pm\rightarrow D \mu^\pm N$. 
\eenum
In both of these production mechanisms, $N$ is produced in association with (quasi-)prompt muons and other objects that can be exploited to discriminate signal events from SM backgrounds. We show these production modes in Fig.~\ref{fig:N_prod}. In our analyses, we focus on $N$ production from $W^\pm$ decay, but comment on the prospects for production from $B^\pm$ decay in ATLAS and CMS in Section \ref{sec:concl}.\\

\begin{figure}[t]
\centering
\includegraphics[width=0.48 \textwidth ]{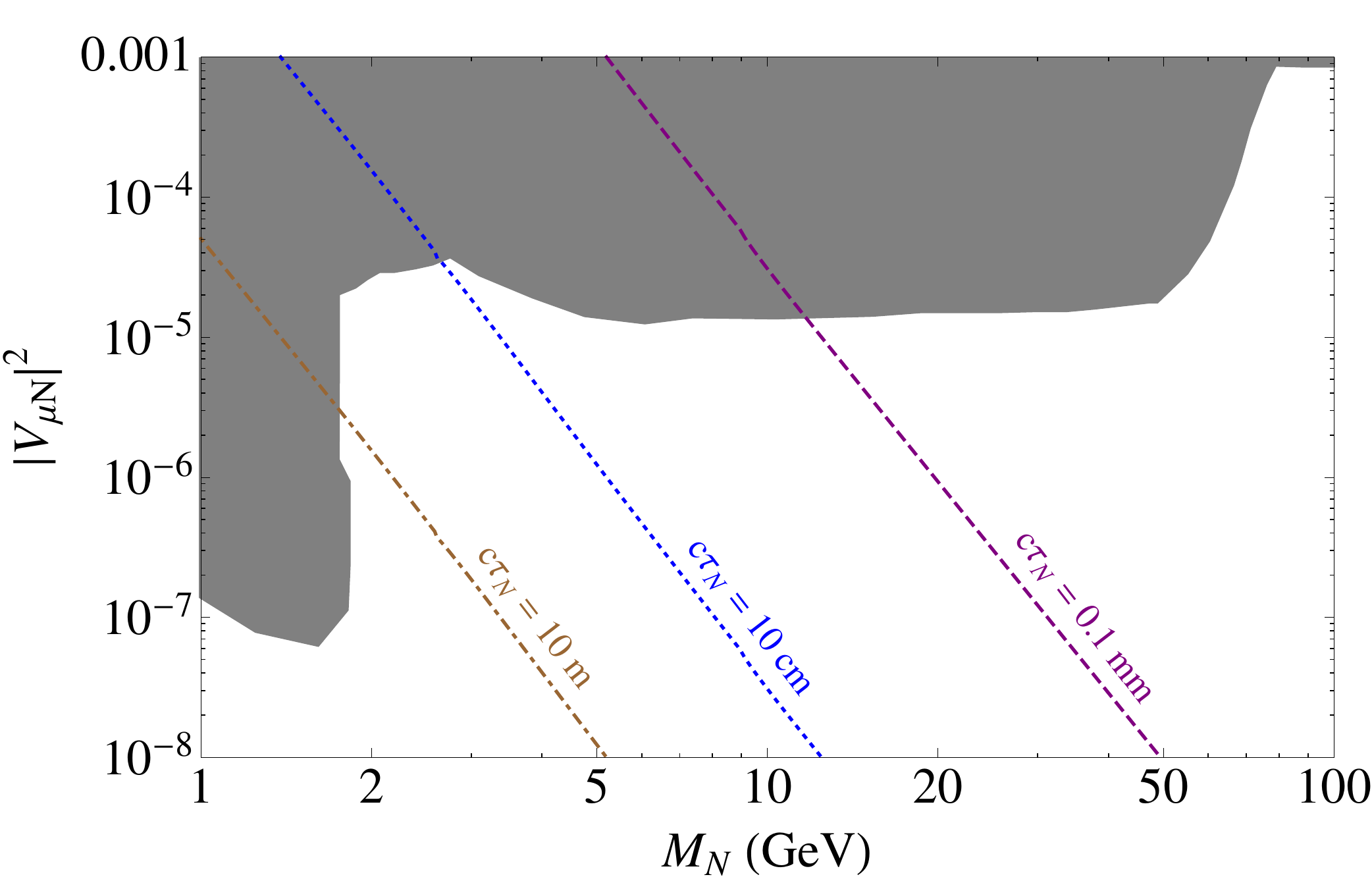}
\caption{Contours of fixed right-handed muon-neutrino proper decay distance, $c\tau_N$, as a function of mixing angle and mass. The contours are:~(right, dashed) $c\tau_N=0.1$ mm; (center, dotted) $c\tau_N=10$ cm; (left, dot-dashed) $c\tau_N=10$ m. The shaded regions are excluded. }
\label{fig:lifetimes}
\end{figure}

\noindent{\bf Decay:} $N$ decays via off-shell $W^\pm/Z$ bosons to other SM states. For the masses we consider, $m_c <  M_N <  M_W$, the RH neutrinos have masses sufficiently above the Quantum Chromodynamics (QCD) confinement scale that their decay rates can be computed in terms of quark and lepton final states. The proper lifetime, $c\tau_N$, as a function of $M_N$ and $V_{\mu N}$ is shown in Fig.~\ref{fig:lifetimes}. Both prompt and displaced decays are expected for RH neutrinos, depending on the mass and mixing angle.

 The most promising decays for the detection of $N$ are:
\benum
\item Leptonic decays, $N\rightarrow \mu^\pm \ell^\mp \nu_\ell$ and $N\rightarrow \nu_\mu \ell^+\ell^-$. In this decay mode, $N$ gives a distinctive final state of multiple charged leptons, potentially of different flavors, and missing energy;

\item Semileptonic decays, $N\rightarrow \mu^\pm q\bar q'$. In this case, $N$ decays are fully visible, allowing a reconstruction of $M_N$ from the decay products, but the hadronic backgrounds are potentially larger. For $M_N<M_W$, there is also typically insufficient transverse momentum in each object to separately reconstruct each of the quarks as jets as well as the charged lepton at the LHC.

\eenum
We find that searches for the clean, fully leptonic signatures to be very sensitive to $N$ production at the LHC for $M_N<M_W$. We consider only decays to electrons and muons. Neglecting charged lepton masses and assuming $M_N \ll M_W$, the partial widths for leptonic decays relevant for our analyses are \cite{Gorbunov:2007ak}
\bea
\Gamma (N\rightarrow \ell_{\alpha}^- \ell_{\beta}^+ \nu_\beta)&=& \frac{ G_{\rm F}^2 M_N^5 | V_{\alpha N} | ^2}{192 \pi^3}\,\,\,\,\,\,(\alpha\neq\beta),\\
\Gamma (N\rightarrow \ell_{\alpha}^- \ell_{\alpha}^+ \nu_\alpha)&=& \frac{ G_{\rm F}^2 M_N^5 | V_{\alpha N} | ^2}{768 \pi^3}\\
&&\times\left(1+4\sin^2\theta_{\rm W}+8\sin^4\theta_{\rm W}\right).\nonumber
\eea
 where $G_{\rm F}$ is the Fermi constant and $\theta_{\rm W}$ is the Weinberg angle. While we show this result for illustrative purposes, in Sections \ref{sec:lepton_jets} and \ref{sec:prompt}, we use the full mass-dependent widths computed numerically in \texttt{Madgraph 5} \cite{Alwall:2014hca}. \\

\noindent{\bf Search Strategies:} In this article, we propose two concrete search strategies to discover $N$, depending on the mass and mixing angle of the RH sterile neutrino.
\begin{enumerate}
\item \emph{Displaced lepton jet:}~For masses $m_c < M_N \ll M_W$, the RH neutrino is boosted in the decay $W^\pm \rightarrow \mu^\pm N$. Therefore, the decay products of $N$ are collimated. In particular, with $N\rightarrow \mu^\pm \ell^\mp \nu$, the $N$ decay products form a lepton jet. The collider signature is a prompt lepton + a single displaced lepton jet, which we discuss in Section \ref{sec:lepton_jets}.

\item \emph{Prompt trileptons:}~For masses $M_N\gtrsim15$ GeV, the RH neutrino is typically neither boosted nor substantially displaced.  The best prospects are in the $W^\pm \rightarrow \mu^\pm N\rightarrow \mu^\pm \mu^\pm e^\mp \nu$ final state, which benefits from smaller SM backgrounds due to the lack of an opposite-sign, same-flavor lepton pair. While others have considered trilepton probes of very high mass RH neutrinos, $M_N\gtrsim 100$ GeV \cite{delAguila:2008cj,delAguila:2008hw,delAguila:2009bb,Chen:2011hc,Das:2012ze,Das:2014jxa,Bambhaniya:2014kga,Dermisek:2014qca}, our proposal in Section \ref{sec:prompt} is, to our knowledge, the first for the key $M_N < M_W$ region of parameter space, showing trilepton searches to be potentially more powerful than the current CMS semileptonic searches \cite{Khachatryan:2015gha}.

\end{enumerate}
In Section \ref{sec:concl}, we also briefly consider the prospects for discovering $N$ produced in $B$ meson decays for masses $M_N\lesssim m_b$.

%% file: lepton_jet.tex
This section focuses on the low-mass, $M_N\ll M_W$ regime. In this mass range the $N$  will typically decay at a length $c\tau \gsim 1 $ mm for masses above a few GeV but less than $\lsim 20\,\,\GeV$, and mixing angles near the current limit $| V_{\mu N} | ^2 \lsim 10^{-5}$, as discussed in Sec.~\ref{sec:simplified_model} and shown in Fig.~\ref{fig:lifetimes}. Consequently the $N$ decay products will not register as prompt final states in the LHC detectors.  As long as the decay occurs within the fiducial volume of either ATLAS or CMS, however, $N$ could still be discovered by looking for displaced decay products. 

There are two possibilities to consider, and each motivates a different search strategy, but we focus on the first of these (see below). We conservatively consider only the case where all of the leptons in the final state are muons, as this scenario is less likely to be contaminated by reducible fake SM backgrounds than when an electron is present in the final state.

\begin{enumerate}
\item $W^\pm\rightarrow \mu^\pm (N\rightarrow \mu^+ \mu^-\nu_\mu)$, which gives a prompt $\mu\, \&\,\mathrm{displaced} ~N \rightarrow 2\mu + \slashed{E_T}$ signature;

\item $W^\pm\rightarrow \mu^\pm (N\rightarrow \mu^\pm q \bar q')$, which gives a prompt $\mu \,\&\, \mathrm{displaced}~N \rightarrow \mu +$ hadronic tracks signature.

\end{enumerate}
In either scenario above, the prompt $\mu$ originates from the direct decay of the parent $W^{\pm}$ boson, and can be used to trigger on the event. Of these two final states, the displaced muon and hadronic tracks signature benefits from a higher branching ratio of $\rm Br$$(W\rightarrow \rm hadrons)$/$\mathrm{Br}(W\rightarrow \mu \nu_\mu)\approx 6$. However,  such a signal region could suffer from potential difficulties, including the possibility that rare displaced hadron decays,  and pile-up-originated tracks can mimic the signal. This background is found to be negligible with sufficiently high track multiplicity and stringent cuts on track $p_{\rm T}$ and vertex mass \cite{ATLAS-CONF-2013-092}, but for $M_N\ll M_W$, the signal tracks typically do not  have enough momentum to pass these cuts. This channel may be viable with relaxed kinematic criteria, but we lack the tools to properly simulate the backgrounds, as ultimately such a signature would be performed using a data-driven estimation of SM backgrounds.  
Therefore, for the low-mass $M_N$ we focus on the first $N$ decay possibility, namely the prompt $\mu+$ displaced $2\mu+\slashed{E_T}$. The prospects for this final state were also addressed in Ref.~\cite{Helo:2013esa}, but we propose a different search strategy based on the distinct signal kinematics discussed below.

\begin{figure}[t]
\centering
\includegraphics[width=0.48 \textwidth ]{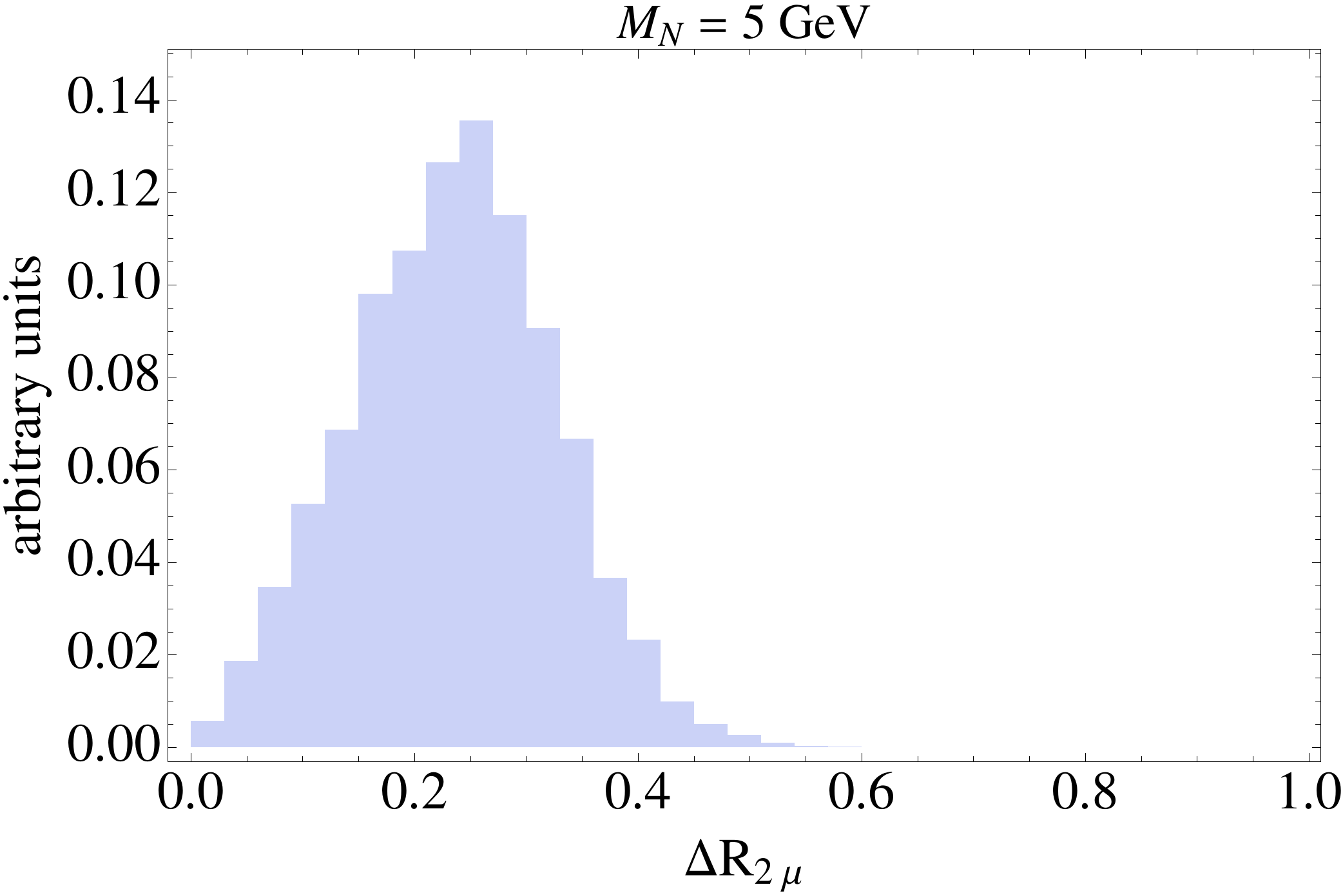}
\includegraphics[width=0.48 \textwidth ]{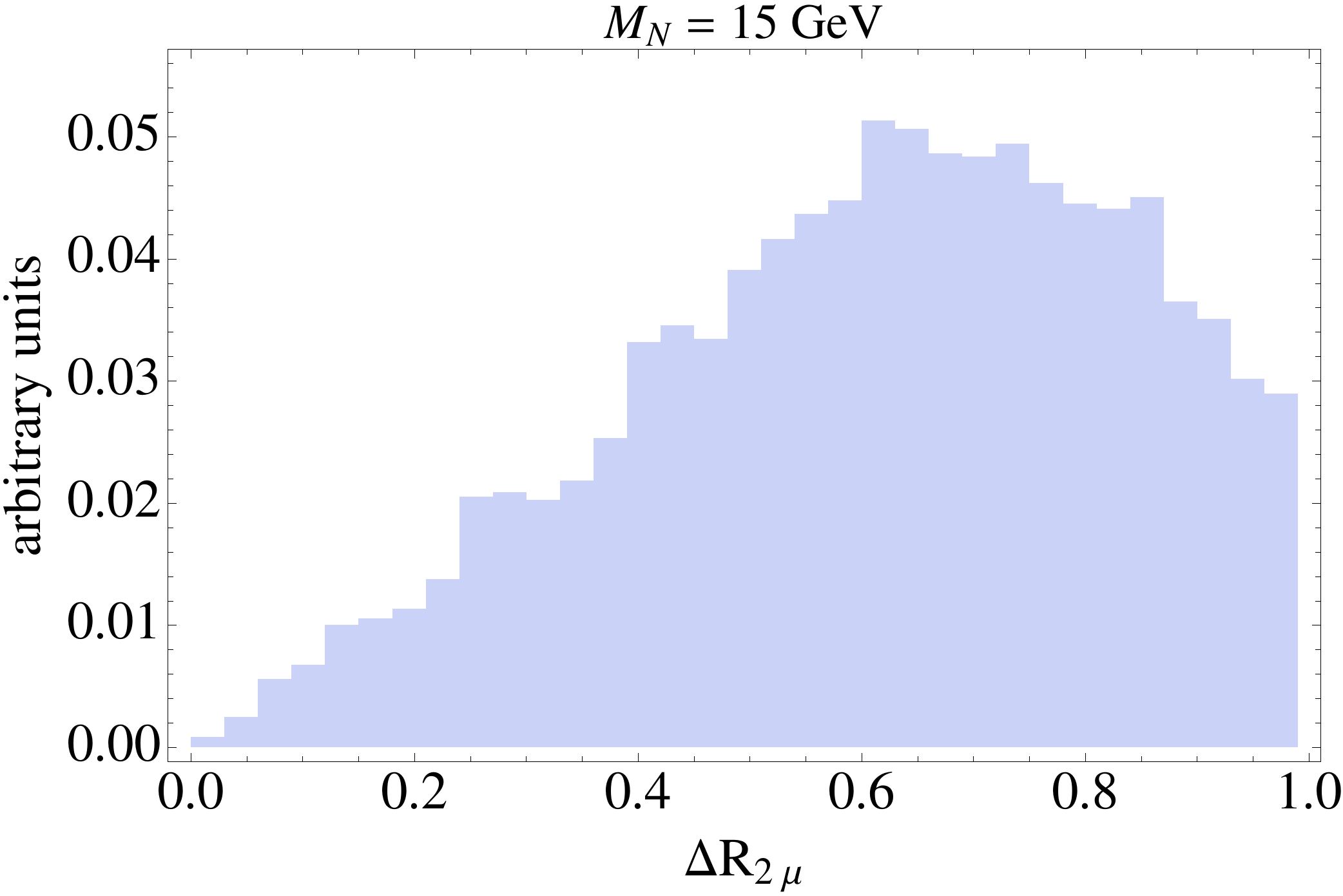}
\caption{The $\Delta R$ separation between the muons originating from the $N\rightarrow \mu^+\mu^-\nu_\mu$ decays for $M_N=5$ GeV (top) and $M_N=15$ GeV (bottom). The $N$ originates from the decay $W^\pm\rightarrow N \mu^\pm$.}
\label{fig:deltaR2mu}
\end{figure}

In the range of $N$ masses and mixing angles giving rise to displaced decays, $M_N\ll M_W$ implies that $N$ are produced with a significant boost from $W$ decay. As a result, the leptons produced in $N$ decays are typically collimated. Fig.~\ref{fig:deltaR2mu} shows the $\Delta R = \sqrt{\Delta \eta^2 + \Delta \phi^2}$ between the two muons from the $N$ decay for $M_N=5$ GeV and $M_N=15$ GeV. For $M_N=5$ GeV the two final-state muons typically appear within less than $\Delta R=0.5$ of one another. A substantial fraction of leptons from $N$ decay for this mass would therefore fail the standard lepton isolation criterion of $\Delta R\lesssim0.3-0.4$. Instead, in analogy with the collimated hadrons in a QCD jet, this signature suggests a \emph{lepton jet} (LJ) - namely a reconstructed object with a more than one leptonic track concentrated within a cone of radius $R_0$. In this study we only address the case of muon-rich final states; thus the object of interest is a \emph{muon jet} ($\mu J$)\footnote{Mixed electron-muon jets are also possible, although electrons are more easily faked by hadron tracks, so we consider the cleaner muon signature to avoid fake backgrounds.}.

We use the definition of a muon jet, $\mu J$, from Ref.~\cite{Aad:2014yea}:
\begin{itemize}
\item {\bf Definition:}~A muon jet is a reconstructed object with \emph{at least} two muons registered in the muon spectrometer (MS) within $\Delta R<R_0=0.5$ of each other.
\end{itemize}

\begin{figure}[t]
\centering
\includegraphics[width=0.48 \textwidth ]{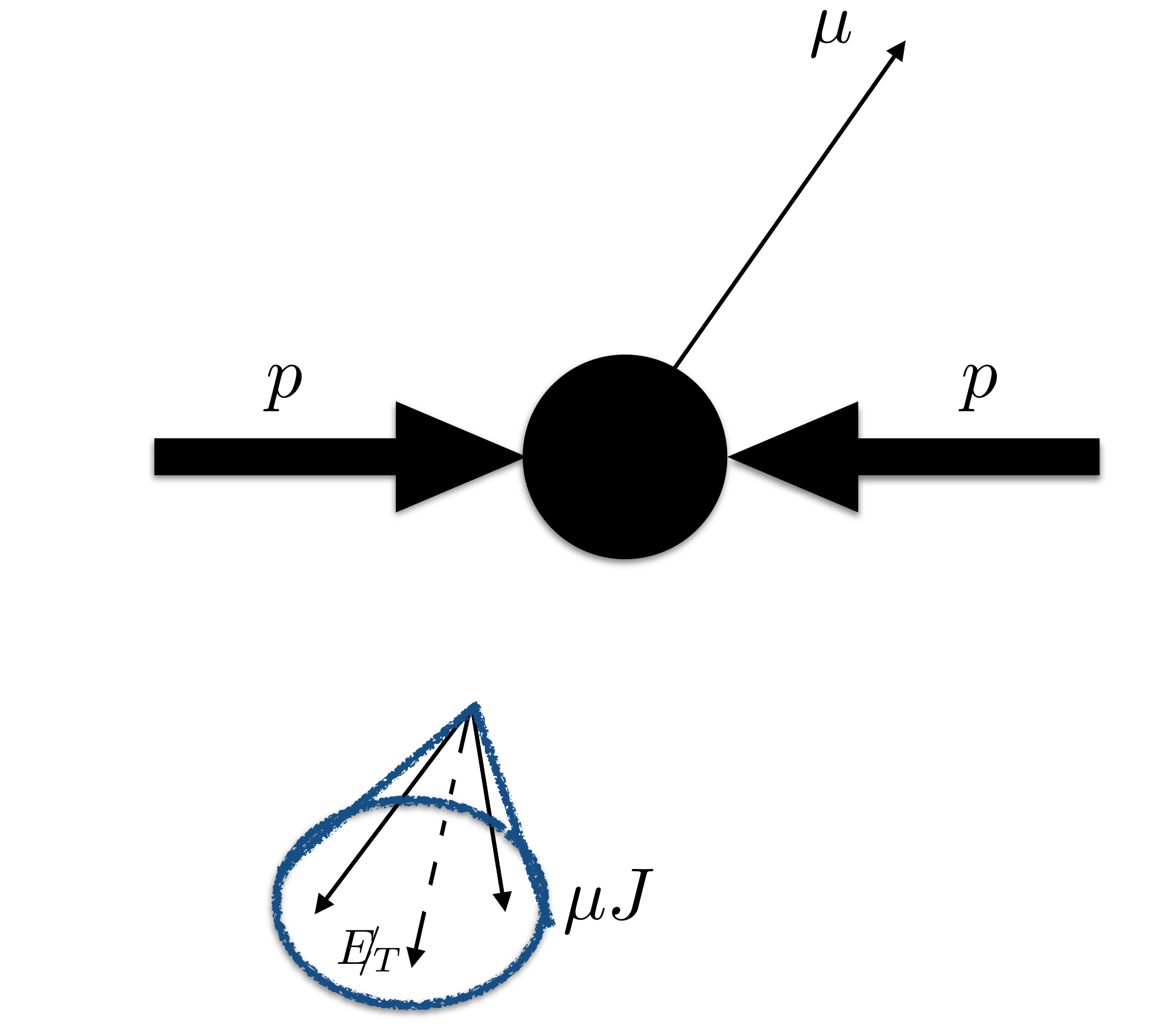}
\caption{An illustration of the final state strategy advocated for the low-mass $N$ regime. The $N$ are produced in the decays of $W^\pm\rightarrow N \mu^\pm$, with the subsequent displaced decay $N\rightarrow \mu^+ \mu^- \nu_\mu$. The muons from the boosted, displaced $N$ will appear as a muon jet $\mu J$ instead of as isolated leptons.}
\label{fig:lowmasscartoon}
\end{figure}

Thus the final state that we study in this section is that of a prompt muon from the $W$ decay + displaced $\mu J$ from $N$ decay. Fig.~\ref{fig:lowmasscartoon} gives a pictorial representation of this signature. While existing lepton jet analyses do exist, their signal regions fail to constrain this part of parameter space as they require two LJs in the event \cite{Aad:2014yea}, whereas our signal gives one per event.

Our $\mu J$ selections are inspired by Ref.~\cite{Aad:2014yea}. The kinematic requirements we use to define the signal region for the low  $M_N$ displaced lepton jet analysis are related:
\begin{enumerate}
\item Trigger on a single muon, requiring a prompt muon track with $p_{\rm T} > 24$ GeV;

\item Any remaining muon tracks must have at least $p_T > 6$ GeV;

\item Demand exactly one $\mu J$, as defined above. The tracks in the $\mu J$ must each have a transverse impact parameter $|d_0|$ between 1 mm and 1.2 m, and the distance between the decay vertex and the primary vertex must be also between 1 mm and 1.2 m  ({\it i.e.,} displaced decay within the tracker);

\item $\Delta R$ between the prompt muon and each of the other muon tracks is less than $\pi - 0.2$ (to veto cosmic muons reconstructed as back-to-back muons decaying in the tracker);

\item Veto significant hadronic activity.

\end{enumerate}

The signal region described above could still be contaminated by cosmic-ray (CR) initiated muon bundles \cite{Aad:2014yea} coinciding with a prompt lepton from a $pp$ collision. We attempt an estimate of the CR background as follows. We use the ATLAS analysis, Ref.~\cite{Aad:2014yea}, which defines a signal region as consisting of at least \emph{two} $\mu J$s in the event. In particular, we focus on their validation region, which uses a trigger on empty bunch crossings; the events seen in that validation sample can only arise from CRs. Assuming that the probability to find a lepton jet from a cosmic ray factorizes\footnote{Due to the fact that a single CR muon is often reconstructed as two back-to-back muons at a displaced vertex, it is likely that the correlation between finding 1 LJ and 2 LJ is very high, making our analysis overly conservative.}, the probability to find one CR is $P(1 \mu J) = \sqrt{N_{2\,\mu J\,\,\mathrm{observed}}/N_{3\mu\,\mathrm{trigger}}} \approx 6\times10^{-3}$ and the LJ rate is $R_{LJ} = P(1 \mu J) \times N_{\rm triggered}/T_{CR}$. Here $T_{CR}\approx (6/15)\times(2/3)\,\,\mathrm{yr}$ is the approximate livetime of the CR data sample acquisition. Since our analysis uses a single muon trigger, we rescale $R_{LJ}$ by the ratio of $R(1\mu\,\mathrm{trigger})/R(3\mu\,\mathrm{trigger})\approx180$ \cite{Hofestaedt:Thesis:2012}, which gives $R_{\rm LJ}\approx 0.018\,\,\mathrm{Hz}$.    We can now use this probability to estimate the number of events in a given livetime period $T$ passing a single prompt muon trigger (with $p_T > 24$ GeV) with a rate $R$; this prompt muon must coincide in a time window $\Delta T$ with a CR-initiated $\mu J$. We then find:

\be
N (1\mu + 1\mu J) &=& R\times T \times (R_{LJ}\times \Delta T)\\
&\approx& 4~\rm{events}
\ee
 for a 3-year livetime, $\Delta T=25$ ns, and assuming an event-filter level trigger rate of $R=100$ Hz \cite{ATLAS:2012yna}.
 
 The estimate above should be taken as a heuristic argument for the CR-initiated backgrounds being a small contamination to the signal region. The CR background could be further reduced by using the high granularity of the tracker and requiring that the two muon tracks within the $\mu J$ reconstruct to the same vertex (which was not required in Ref.~\cite{Aad:2014yea}).  Kinematic features such as the invariant mass of the $\mu J$ and the alignment of $\slashed{E_T}$ with the $\mu J$ could be used to further suppress backgrounds. Therefore, we assume a background-free search with integrated luminosity of $300\,\,\mathrm{fb}^{-1}$, and define our $2\sigma$ exclusion reach contours by requiring 3 signal events after cuts.

\begin{figure}[t]
\centering
\includegraphics[width=0.48 \textwidth ]{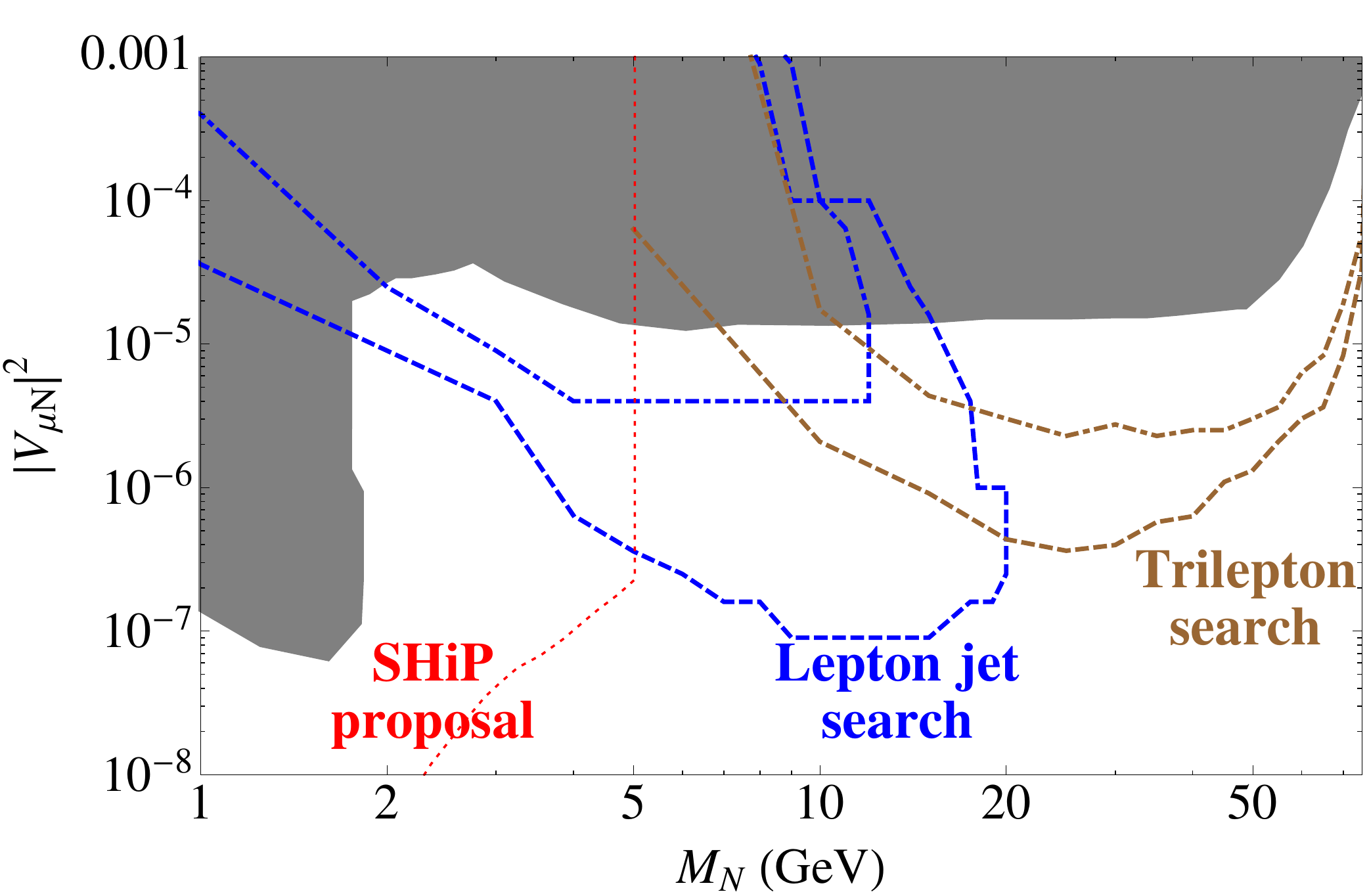}
\caption{95\% confidence level reach of our proposed lepton jet and trilepton searches in terms of the sterile neutrino simplified model parameters. The blue lines show the reach of the displaced lepton jet search at (dot-dashed) $\sqrt{s}=8$ TeV with $20$ fb$^{-1}$, (dashed)  $\sqrt{s}=13$ TeV with $300\,\,\mathrm{fb}^{-1}$. The brown lines show the prompt trilepton reach with (dot-dashed) $\sqrt{s}=8$ TeV with $20\,\,\mathrm{fb}^{-1}$ and 50\% systematic uncertainty, (dashed) $\sqrt{s}=13$ TeV with $300\,\,\mathrm{fb}^{-1}$ and 20\% systematic uncertainty. The thin red dotted line shows the reach for the proposed SHiP experiment from Ref.~\cite{Deppisch:2015qwa}. The shaded region is excluded. }
\label{fig:prompt_reach}
\end{figure}
 
We perform the simulation for the low-mass $N$ signal region using \texttt{Madgraph 5} \cite{Alwall:2014hca}. Because of the all-muon final state, we consider only parton-level events. We show our estimated sensitivity at the LHC for this signal region in Fig.~\ref{fig:prompt_reach}, both for 8 TeV with 20 fb$^{-1}$ and for 13 TeV with 300 fb$^{-1}$. For masses near $M_N\approx 15$ GeV, the sensitivity of this analysis could be further improved by increasing the $\Delta R_0$ in the definition of  $\mu J$ as the $N$ decay products' separation increases. Furthermore, the requirement that the $\mu J$ appear at a displaced vertex in the tracker ($|d_0| \lesssim 1$m) could also be relaxed to consider DVs in the calorimeters and the muon spectrometer, but  the background estimate from Ref.~\cite{Aad:2014yea} has to be modified for this case. 

%% file: prompt.tex
For masses $M_N\gtrsim 15$ GeV, $N$ typically decays promptly, and the reconstruction of the decay vertex no longer provides significant discriminating power from SM backgrounds. In this section, we investigate the most promising final states for discovering $N$ in the prompt regime. In particular, we find that targeted searches in the trilepton final state with no opposite-sign, same-flavor (OSSF) leptons can suppress SM backgrounds and give a smoking gun signature for lepton-number-violating RH neutrinos with $M_N\lesssim M_W$. While trilepton final states have been considered previously for $M_N\gtrsim M_W$ and/or Dirac neutrinos \cite{delAguila:2008cj,delAguila:2008hw,delAguila:2009bb,Chen:2011hc,Das:2012ze,Das:2014jxa,Bambhaniya:2014kga}, we show that the $M_N\lesssim M_W$ regime presents the LHC experiments with different kinematics than previously considered. By tailoring the signal selection to the softer kinematic regime, we show that trilepton searches have the capability of probing Majorana $N$ down to $M_N\sim 10$ GeV.

The only current analysis at the LHC for $N$ in the $M_N\lesssim M_W$ mass range is a CMS search in the $W^\pm \rightarrow \mu^\pm\mu^\pm jj$ final state \cite{Khachatryan:2015gha}. This search was originally designed for $M_N\gg M_W$ \cite{Keung:1983uu,Pilaftsis:1991ug,Datta:1993nm}, and has recently been re-optimized for $M_N\lesssim M_W$ \cite{Khachatryan:2015gha}. The re-analysis requires two same-sign muons with $p_{\rm T}>15$ GeV and two jets with $p_{\rm T}>20$ GeV, and seeks to reconstruct $M_{\mu^\pm\mu^\pm j j}\sim M_W$. It is immediately obvious that, for $N$ produced in the decay of $W^\pm$, there is insufficient phase space to pass all of the kinematic cuts unless the $W^\pm$ is highly boosted; however, if the $W^\pm$ is boosted, the jets in the decay of $N$ are not separately resolved. Therefore, this search suffers from extremely tiny signal efficiencies for $M_N<M_W$ ($\sim0.6-0.8\%$), and for signal events passing all cuts, one of the jets is typically not from the $N$ decay. This can be deduced from the fact that $M_{\mu^\pm\mu^\pm j j}$ peaks \emph{well above} $M_W$ for the signal in Ref.~\cite{Khachatryan:2015gha}, whereas the correctly reconstructed decay products of $N$ should always give a mass \emph{below} $M_W$. This suggests that, even for signal events, one of the final-state jets is uncorrelated with the $N$ decay products, and so the (small) background looks nearly identical to the signal. Thus, the constraints from the $\mu^\pm\mu^\pm j j$ search are only comparable to or worse than the LEP constraints for $M_N\lesssim M_W$. 

\begin{figure}[t]
\centering
\includegraphics[width=0.35 \textwidth ]{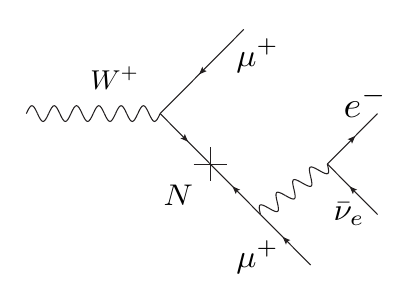}
\caption{Production and decay of $N$ for our proposed prompt trilepton search with no OSSF lepton pairs.}
\label{fig:feynman_prompt}
\end{figure}

Given the challenges in reconstructing both quarks from $N\rightarrow \mu^\pm q \bar q'$ decay as separate jets, we consider instead the purely leptonic decay, $W^\pm \rightarrow \mu^\pm N \rightarrow 3\ell + \nu$. We  propose exploiting the Majorana nature of the sterile neutrino to look for $W^\pm \rightarrow \mu^\pm N\rightarrow \mu^\pm \mu^\pm e^\mp \nu_e$ final states (see Fig.~\ref{fig:feynman_prompt}):~because there are no OSSF lepton pairs in the final state, SM backgrounds involving $\gamma^*/Z$ are greatly suppressed.

Current experimental searches in trilepton final states have targeted supersymmetric final states with large $\cancel{E}_{\rm T}$, although CMS has an analysis with low $\cancel{E}_{\rm T}$ and low $H_{\rm T}$ \cite{Chatrchyan:2014aea}. This search has been recast for $M_N>M_W$ \cite{Das:2014jxa}, and here we recast the analysis to determine the constraints on the low-mass signal region $M_N\lesssim M_W$. In particular, we use the OSSF-0 signal region to find the most powerful bound. Using the data from the $\cancel{E}_{\rm T}<50$ GeV, $H_{\rm T}<200$ GeV, OSSF-0 bin with 0 $b$-jets from Ref.~\cite{Chatrchyan:2014aea}, we find that the CMS trilepton analysis is competitive with, but does not quite surpass, the LEP and CMS $\mu^\pm\mu^\pm jj$ analyses for $20\,\,\mathrm{GeV}\lesssim M_N \lesssim M_W$. Given, however, that a non-optimized analysis in the OSSF-0 trilepton channel already gives a bound competitive with other search channels, this suggests that a targeted search for $N$ in the trilepton final state would give a significant improvement in sensitivity. \\

\noindent {\bf Monte Carlo simulations:}~We perform a Monte Carlo (MC) analysis to estimate the improvement in sensitivity that can be obtained with a targeted trilepton search for sterile neutrinos. We simulate parton-level processes in \texttt{Madgraph 5} \cite{Alwall:2014hca} and shower the events in \texttt{Pythia 6} \cite{Sjostrand:2006za}. For background processes, matrix elements with up to two extra partons are simulated and matched to the shower using the MLM-based shower-$k_\perp$ scheme \cite{Alwall:2008qv}. The dominant backgrounds are $\gamma^*/Z+$ jets, $t\bar t$, and $WZ+$ jets. Jets are clustered using the anti-$k_{\rm T}$ algorithm \cite{Cacciari:2008gp} implemented with the \texttt{Fastjet 3} package \cite{Cacciari:2011ma}. Signal and background cross sections are normalized to their next-to-leading-order values \cite{Campbell:2011bn,Campbell:2013yla,stirlingxsec}.

A major obstruction to background simulation is that the dominant backgrounds for OSSF-0, low-$\cancel{E}_{\rm T}$ and low-$H_{\rm T}$ trilepton searches come from processes where one or more ``non-prompt'' (fake) leptons are present in the final state. For example, $Z/\gamma^*+$ jets and $t\bar t$ backgrounds can fake trilepton signatures if one of the final-state jets is mis-tagged as a lepton; this ``fake'' can either come from an actual lepton originating from a heavy-flavor meson decay, or from light hadrons that are mis-reconstructed as leptons. Because fake leptons are very rare and may rely on improperly modelled jet fragmentations, MC estimates for fake leptons are unreliable, and
the ATLAS and CMS collaborations instead use a data-driven approach to estimate lepton fakes in their multilepton analyses \cite{Chatrchyan:2014aea,ATLAS:2014kca,Khachatryan:2015gha}. Since we do not have access to the resources needed for data-driven estimates, we adopt an approach proposed by Ref.~\cite{Curtin:2013zua}, which takes jet-enriched samples and constructs a map from jet kinematics to fake lepton kinematics. This method allows for the use of reasonably sized samples of $Z/\gamma^*+$ jets and $t\bar t$ events to obtain sufficient statistics for estimating fake lepton backgrounds. We describe the procedure and validation of this method in Appendix \ref{app:fakesim}. \\

\noindent {\bf Signal kinematics:}~We apply basic selection criteria similar to the OSSF-0, 0 $b$-jet bin for the CMS trilepton analysis \cite{Chatrchyan:2014aea}. Requiring a leading lepton with $p_{\rm T}>20$ GeV and all subleading leptons with $p_{\rm T}>10$ GeV, we demand exactly three leptons, zero OSSF lepton pairs, and zero $b$-tagged jets with $p_{\rm T}>30$ GeV (using the b-tagging working point from Ref.~\cite{Chatrchyan:2014aea}). Defining $H_{\rm T}$ as the scalar $p_{\rm T}$ sum of jets with $p_{\rm T}>20$ GeV, we apply upper cuts on $H_{\rm T}$ and $\cancel{E}_{\rm T}$ to suppress $t\bar t$ backgrounds.  For the histograms shown below, we apply $H_{\rm T}<200$ GeV and $\cancel{E}_{\rm T}<50$ GeV, although stricter cuts are applied for the final analysis.

\begin{figure}[t]
\centering
\includegraphics[width=0.48 \textwidth ]{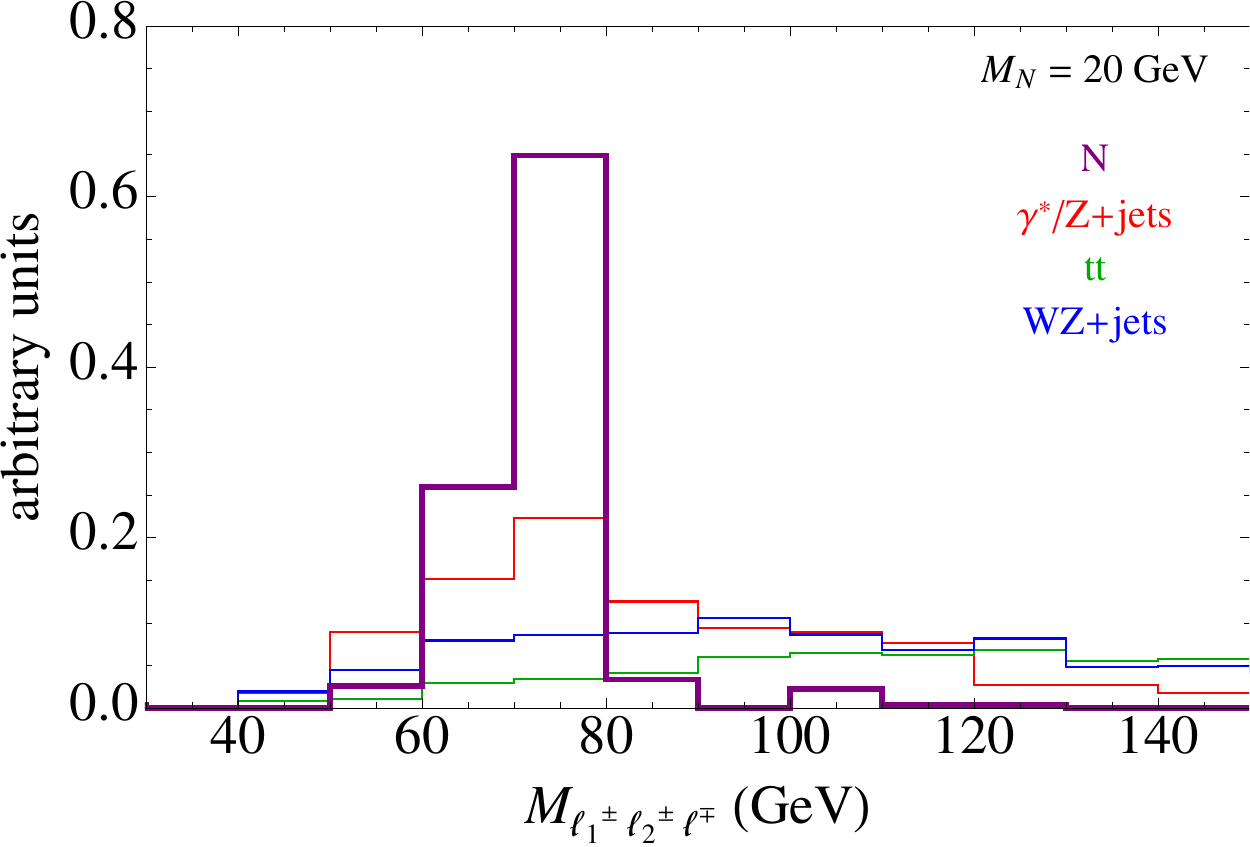}
\includegraphics[width=0.48 \textwidth ]{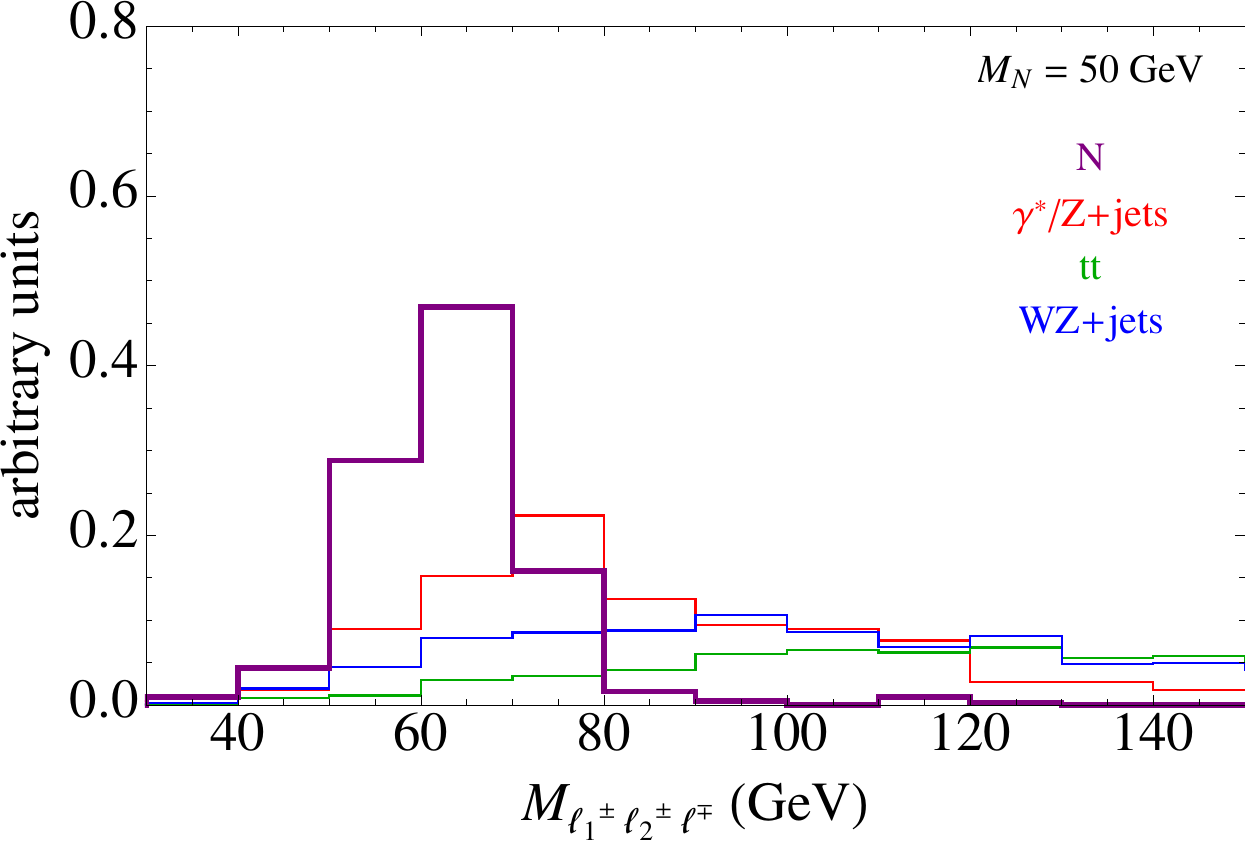}
\caption{Histograms of $M_{\ell_1^\pm\ell_2^\pm\ell^\mp}$ in the OSSF-0, 0-$b$, $H_{\rm T}<200$ GeV, $\cancel{E}_{\rm T}<50$ GeV bin. For both signal mass points, there is a cutoff in the distribution at $M_W$, with the peak more prominent for smaller $M_N$.}
\label{fig:m3l}
\end{figure}

Taking as our convention that  $\ell_1^\pm $ ($\ell_2^\pm$) is the hardest (softest) same-sign lepton, and $\ell^\mp$ is the lepton of opposite sign, we study various kinematic distributions of the charged leptons. In particular,  we find two observables that are powerful discriminants between signal and background. The first is the trilepton invariant mass, $M_{\ell_1^\pm \ell_2^\pm \ell^\mp}$; because the invariant mass of the three leptons plus the neutrino reconstructs the $W$, this distribution has a sharp cutoff at $M_W$, as we show in Fig.~\ref{fig:m3l}\footnote{In the CMS search for semileptonic decays of $N\rightarrow \mu^\pm \mu^\pm jj$ \cite{Khachatryan:2015gha}, this type of kinematic observable is used, but because one of the jets is typically from uncorrelated radiation in signal events, the signal peaks above $M_W$ and the observable is not an effective discriminant of signal from background. This is in contrast with our findings for trilepton searches.}. Indeed, for masses $M_N\ll M_W$, most of the energy from the $W$ decay must go into the charged leptons for a high efficiency of passing kinematic cuts, and therefore a strong peak is observed even though one of the final states is invisible. For $M_N\sim M_W$, all of the leptons (including the neutrino) arising from the $N$ decay are relatively hard, while the muon from the original $W$ decay is soft in turn, and so the $W$ peak is not as pronounced.

\begin{figure}[t]
\centering
\includegraphics[width=0.48 \textwidth ]{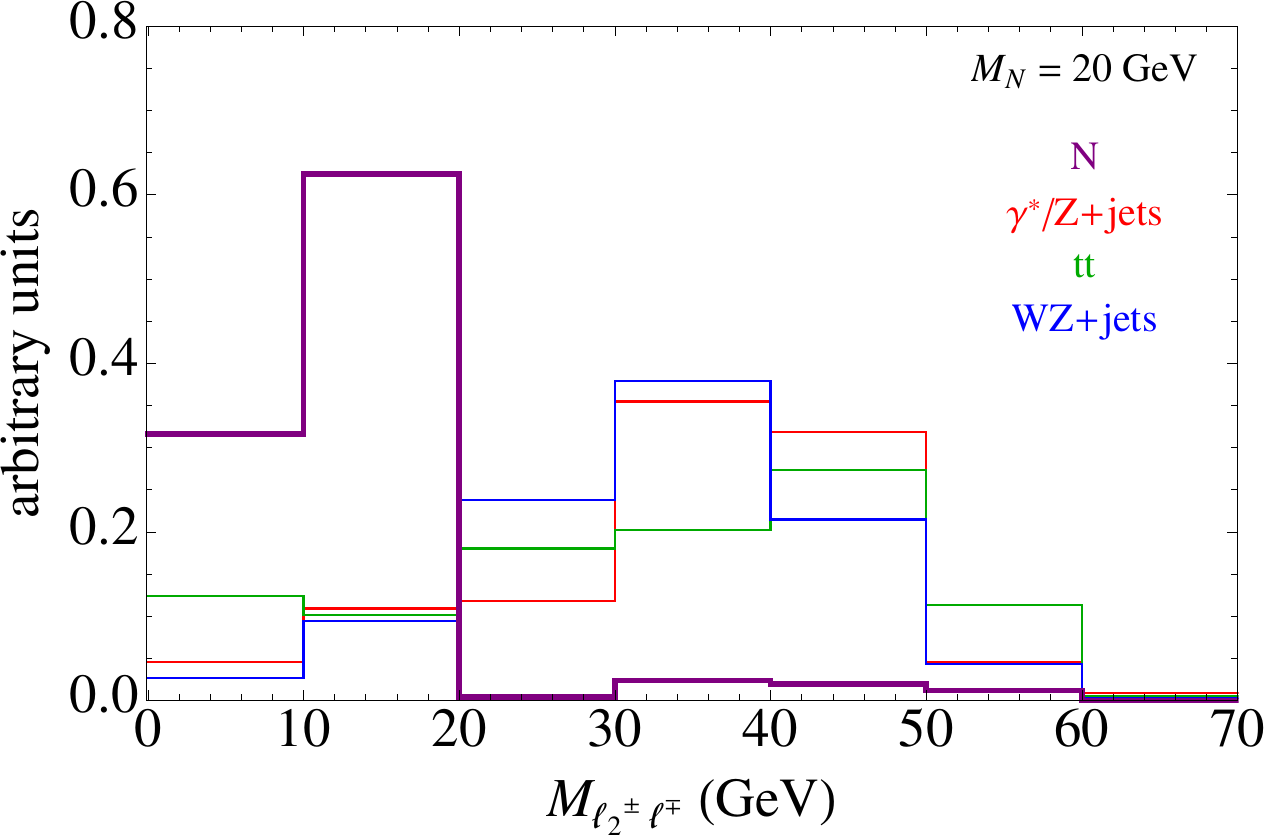}
\includegraphics[width=0.48 \textwidth ]{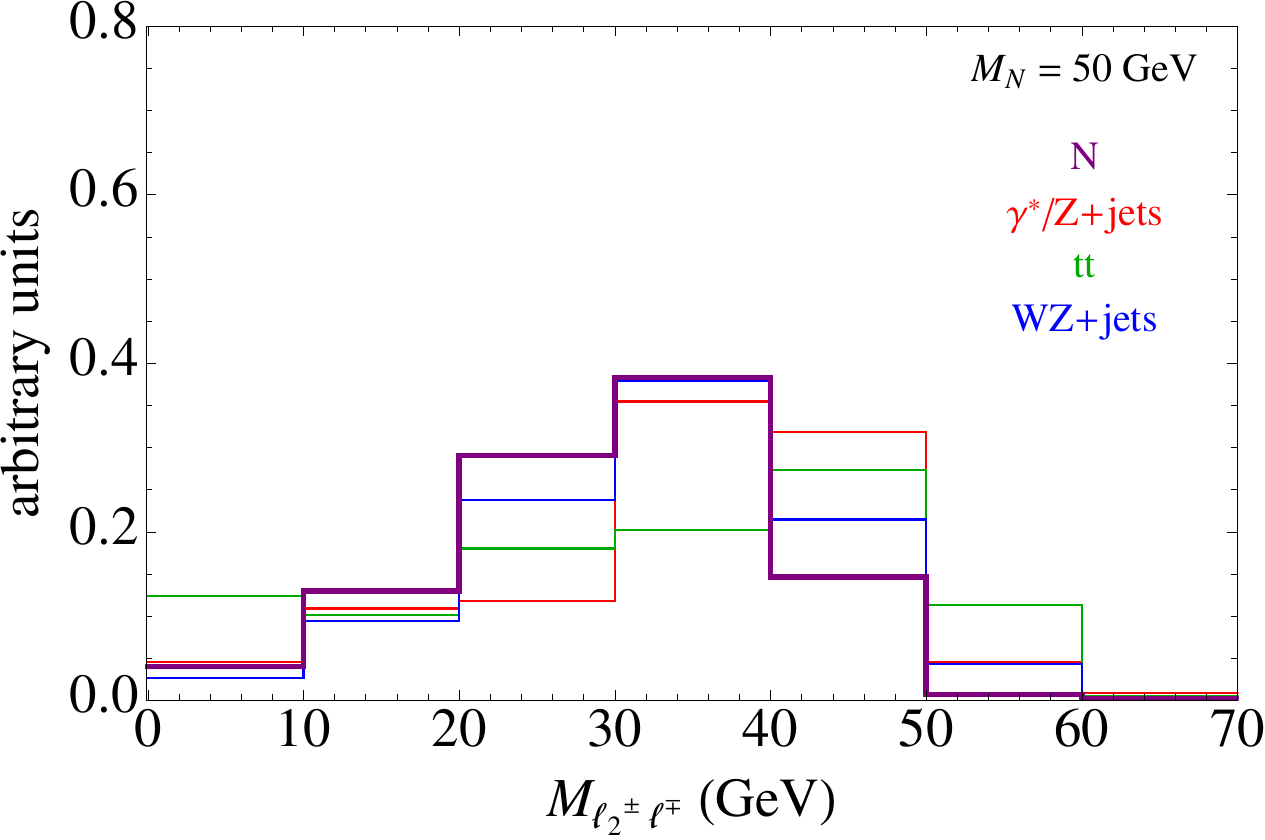}
\caption{Histograms of $M_{\ell_2^\pm\ell^\mp}$ in the OSSF-0, 0-$b$, $H_{\rm T}<200$ GeV, $\cancel{E}_{\rm T}<50$ GeV, $M_{\ell_1^\pm \ell_2^\pm \ell^\mp}<80$ GeV bin. For both signal mass points, the signal peaks below $M_N$, with the peak more prominent for smaller $M_N$.}
\label{fig:m2l}
\end{figure}

The second useful discriminant  is the invariant mass $M_{\ell_2^\pm \ell^\mp}$. To clearly show its effects, we impose a cut of $M_{\ell_1^\pm \ell_2^\pm \ell^\mp}<80$ GeV in addition to the selection criteria described above. For $M_N\ll M_W$, $\ell_2^\pm$ and $\ell^\mp$ both come from the decay of $N$, and so their invariant mass is kinematically restricted to be $<M_N$, which we show in Fig.~\ref{fig:m2l}. For $M_N\ll M_W$, a peak structure is again obtained, while for $M_N\sim M_W$, all three charged leptons have comparable momenta, and it is no longer possible to identify which lepton came from the $N$ decay. Thus, the $M_{\ell_2^\pm \ell^\mp}$ distribution is broader and looks more like background, degrading the discriminating power (as shown in Fig.~\ref{fig:m2l} with $M_N=50$ GeV). \\

\noindent {\bf Proposed analysis:}~Taking into account all of the above, we impose the following kinematic selections:
\begin{enumerate}
\item Two prompt isolated same-sign muons and one prompt isolated opposite-sign electron with $p_{\rm T}>10$ GeV (leading $p_{\rm T}>20$ GeV);
\item $H_{\rm T}<50$ GeV and $\cancel{E}_{\rm T}<40$ GeV (to further suppress fake $t\bar t$ and $Z\rightarrow \tau\tau$ backgrounds)\footnote{Our simulations indicate that, for signal, $H_{\rm T}$ and $\cancel{E}_{\rm T}$ may peak even lower than this; however, resolution effects, initial state radiation, and pile-up become important for smaller values of these variables, and so the optimal cut value should be determined by ATLAS and CMS.};
\item $M_{\ell_1^\pm\ell_2^\pm\ell^\mp}<80$ GeV (for $M_N<50$ GeV, we additionally require $M_{\ell_1^\pm\ell_2^\pm\ell^\mp}>60$ GeV);
\item An upper cut on $M_{\ell_2^\pm\ell^\mp}$, separately optimized for each value of $M_N$. 
\end{enumerate}
We optimize the cuts on $M_{\ell_2^\pm\ell^\mp}$ for each $M_N$ to maximize signal significance for  $\sqrt{s}=13$ TeV and $300\,\,\mathrm{fb}^{-1}$, scanning over values of the cut ranging from 10-60 GeV. The optimal cut is typically just below $M_N$. We determine the signal reach at 95\% confidence level, with the number of signal events exceeding $\mathrm{max}(3,2\sigma_{\rm bkd})$ for a given integrated luminosity. We include both statistical uncertainty and a projected systematic uncertainty in $\sigma_{\rm bkd}$. 

The projected reach is determined both for $\sqrt{s}=8$ TeV and $20\,\,\mathrm{fb}^{-1}$ (50\% systematic), as well as $\sqrt{s}=13$ TeV and $300\,\,\mathrm{fb}^{-1}$ (20\% systematic)\footnote{In a data-driven background estimate, an increase in statistics in the control region will lead to a smaller uncertainty in the extrapolation into the signal region, which explains our choice of a smaller systematic uncertainty with the larger dataset at 13 TeV.}. For comparison, the current CMS search for Majorana neutrinos in the $2\mu^\pm+2j$ channel is dominated by fakes and the systematic uncertainty on the background estimate is $\sim30\%$ \cite{Khachatryan:2015gha}. We show our estimate of the trilepton search reach  in Fig.~\ref{fig:prompt_reach}. For $10\,\,\mathrm{GeV}\lesssim M_N\lesssim M_W$, the prompt trilepton search can substantially improve upon the LEP bound, and a search with Run I data may already cover new territory. The sensitivity diminishes at low mass because the leptons from $N$ decays originate from a displaced vertex; this region is best covered by the lepton jet search proposed in Section \ref{sec:lepton_jets}. In the overlapping region, performing both searches can improve sensitivity to $N$ and increase confidence in a putative signal. For $M_N\sim M_W$, the reach weakens because the decay $W^\pm \rightarrow \mu^\pm N$ becomes phase-space-suppressed.

In summary, the search for $N$ in prompt trilepton final states nicely complements the proposals for long-lived searches at fixed-target experiments and at the LHC, and is a promising channel for probing RH neutrinos in both Run I and Run II data.

%% file: concl.tex
Theories with right-handed neutrinos at or below the weak scale can resolve many of the shortcomings of the SM, providing an explanation for neutrino masses, the origin of the baryon asymmetry, and a candidate for dark matter. We have shown that, in the well-motivated mass range $m_c\lesssim M_N\lesssim M_W$, the LHC can serve as a powerful probe of right-handed neutrinos due to the large number of $W$ bosons produced, complementing the lower-mass searches at fixed-target probes such as the proposed SHiP experiment. In particular, the leptonic interactions of a Majorana neutrino give rise to distinctive multilepton searches in displaced and/or prompt final states.

We have proposed two search strategies that can improve the reach of right-handed neutrinos by up to two orders of magnitude in cross section:~a search for a displaced lepton jet in association with a prompt lepton, which covers the mass range $m_c\lesssim M_N\lesssim20$ GeV, and a prompt trilepton search that covers the mass range $10\,\,\mathrm{GeV}\lesssim M_N\lesssim M_W$.  Optimizing the kinematic selections with respect to background can lead to clean signatures with good discovery potential, even though the decay products of right-handed neutrinos are relatively soft. Because the final states are entirely leptonic, the dependence on pile-up and other factors with high-energy/luminosity running should not significantly degrade the prospects for the discovery of right-handed neutrinos.

Before concluding, we discuss briefly the prospects for right-handed neutrino searches in the range $M_N\lesssim m_b$ at ATLAS/CMS. Here, the dominant production mechanism is not from $W$ decays, but from $B$-hadron decays such as $B^\pm \rightarrow D \mu^+ N$. Na\"ively, it seems that the strong production cross section of $N$ in this mass range could lead to reasonable discovery prospects. Furthermore, the RH neutrinos are typically very long-lived in this mass range ($c\tau\gtrsim1-10$ m), and so one can take advantage of the large ATLAS/CMS detectors to look for $N$ decays in the muon spectrometer \cite{ATLAS:2012av}. However, we find that for $m_b -m_c \lesssim M_N \lesssim m_b$, the $\tau_N$ results in decays inside the muon spectrometer, but the $B$ is kinematically forced to decay through its mixing with the up quark, and this leads to a suppressed $N$ production rate. Conversely, for $M_N\lesssim m_b - m_c$, the branching fraction of $B\rightarrow N$ is larger, but the lifetime is longer so that one must pay a severe penalty in rate to require the decay length to be $\lesssim10$ m. Combining this with the requirement of having sufficiently hard final states to reconstruct the lepton from $B$ decay and the displaced vertex inside the muon spectrometer, we estimate the projected sensitivity to be worse than the current constraints. Thus, the lepton-jet final state from $W$ decay is still the best way to search for RH neutrinos for $M_N\lesssim m_b$ with ATLAS/CMS, although direct searches of $N$ production from $B$ decays at $B$-factories and LHCb will have sensitivity in this mass range \cite{Liventsev:2013zz,Aaij:2014aba,Canetti:2014dka}.\\

{\bf Acknowledgements:~} We thank Guido Ciapetti, David Curtin, Miriam Diamond, Henry Lubatti, Philippe Mermod, and Ennio Salvioni for helpful discussions. This work was made possible by the facilities of the Shared Hierarchical 
Academic Research Computing Network (SHARCNET) and Compute/Calcul Canada. This research was supported in part by Perimeter Institute for Theoretical Physics. Research at Perimeter Institute is supported by the Government of Canada through Industry Canada and by the Province of Ontario through the Ministry of Research and Innovation. EI is partially supported by the Ministry of Research and Innovation - ERA (Early Research Awards) program.

%% file: fakesim.tex
We use the method of fake lepton simulation proposed in Ref.~\cite{Curtin:2013zua}. This approach relies on the fact that fake leptons are initiated by jets, and the leptons consequently inherit the kinematic properties of the jet. Therefore, fake lepton kinematics can be estimated from jet-rich MC samples by mapping  jet kinematics onto the fake lepton momenta. In particular, the authors of Ref.~\cite{Curtin:2013zua} found that lepton fakes can be well-estimated by applying:~a \emph{mistag rate}, $\epsilon_{j\rightarrow\ell}(p_{\rm T})$, which is a function of the jet $p_{\rm T}$; and a \emph{transfer function}, $\mathcal T_{j\rightarrow\ell}$, which is a probability distribution function that maps $p_{\mathrm{T}j}$ into the fake $p_{\mathrm{T}\ell}$. These have the functional forms:
\bea
\epsilon_{j\rightarrow\ell}(p_{\mathrm{T}j}) &=& \epsilon_{200}\left[1-(1-r_{10})\frac{200 - p_{\mathrm{T}j}/\mathrm{GeV}}{200-10}\right],\\
\mathcal{T}_{j\rightarrow\ell}(\alpha) &=& \frac{1}{\mathcal N} \exp\left[-\frac{(\alpha-\mu)^2}{2\sigma^2}\right]. \label{eq:transfer}
\eea
In this expression, $p_{\mathrm{T}\ell} \equiv (1-\alpha) p_{\mathrm{T}j}$,  $\epsilon_{200}$ parameterizes the overall mistag rate, $r_{10}$ gives the gradient for the $p_{\rm T}$-dependent mistag rate, and $\mu$ and $\sigma$ are the mean and standard deviation of the Gaussian distribution mapping the jet momentum to the lepton momentum. If $\alpha\neq0$, some of the jet momentum is not mapped into the lepton momentum, and it is assumed to contribute to $\cancel{E}_{\rm T}$. The Gaussian function Eq.~(\ref{eq:transfer}) is truncated so that $0\leq \alpha\leq1$, and $\mathcal{N}$ is a normalization factor.

 We find that agreement with experimental data is obtained only if we assume that all fake leptons arise from jets containing heavy-flavor hadrons. Consistently, we find that a good fit to data is obtained with the inputs $r_{10}=1$, $\mu=0.5$, and $\sigma =0.3$. These parameters give a flat mistag rate in jet $p_{\rm T}$, and equally divide the ``jet'' energy between the fake lepton and the neutrino, which is expected from leptons originating in heavy-flavor decays. These parameters are somewhat different from those obtained in Ref.~\cite{Curtin:2013zua}, but in both their analysis and our own, we find that changing the parameters in the efficiency and transfer function does not substantially change the fit, provided the total normalization remains fixed. This is due to approximate degeneracies present among the various parameterizations of the fake lepton functions:~for example, a softer momentum spectrum can be obtained both with a large $\mu$, in which case the lepton momentum is imparted with less of the original jet momentum, or $r_{10}\approx 1$, which gives a larger weighting to soft momenta relative to $r_{10}=0$. In our comparison with the data described below, we find that varying the fake simulator input parameters changes the contributions to various bins by a factor $\lesssim2$, and similarly, the reach in $|V_{\mu N}|^2$ of our trilepton analysis similarly changes by $\lesssim2$.

\begin{figure}[t]
\centering
\includegraphics[width=0.48 \textwidth ]{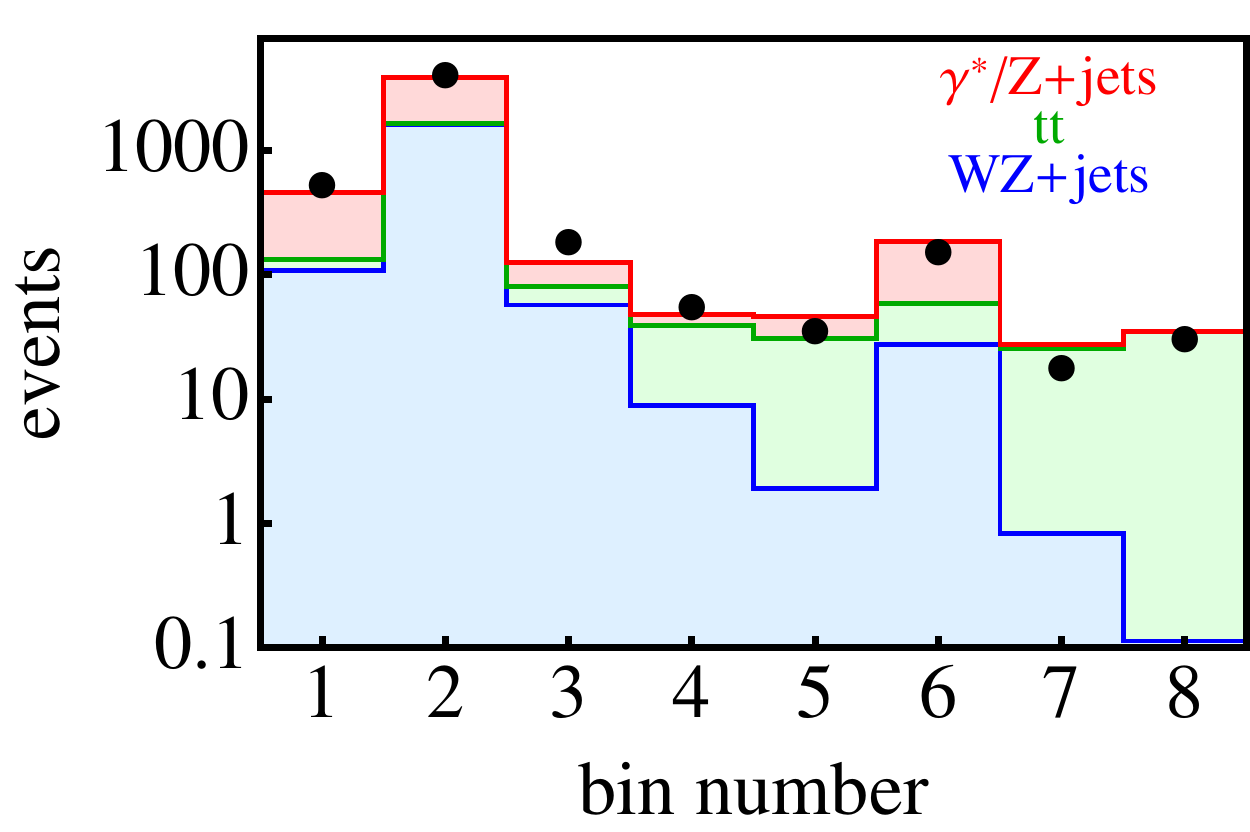}
\caption{Comparison of our MC simulation of $\gamma^*/Z+$ jets (top, red), $t\bar t$ (middle, green), and $WZ+$jets (bottom, blue) with the observed values from the CMS trilepton analysis (black dots) \cite{Chatrchyan:2014aea}. The analysis is at $\sqrt{s}=8$ TeV, $\mathcal{L} =20\,\,\mathrm{fb}^{-1}$, and applies all of the selections in Ref.~\cite{Chatrchyan:2014aea} including the cuts $\cancel{E}_{\rm T}<50$ GeV and $H_{\rm T}<200$ GeV. The bins are divided according to the number of $b$-tagged jets, whether there exists an OSSF lepton pair, and the mass of any OSSF lepton pair. The bins are:~1) 0-$b$, OSSF-1, $M_{\ell^+\ell^-}<75$ GeV; 2) 0-$b$, OSSF-1, $|M_{\ell^+\ell^-}-M_Z|<15$ GeV; 3) 0-$b$, OSSF-1, $M_{\ell^+\ell^-}>105$ GeV; 4) 0-$b$, OSSF-0. Bins 5-8 are the same as bins 1-4, but with at least one $b$-tagged jet.}
\label{fig:trilepton_validation}
\end{figure}

As our paper focuses on trilepton signatures, we fix the overall fake efficiency using the CMS trilepton data \cite{Chatrchyan:2014aea}:~we normalize our background estimates to the OSSF-1, 0 $b$-jet, $M_{\rm \ell^+\ell^-}\sim M_Z$ bin (``bin 2'' in Fig.~\ref{fig:trilepton_validation}), and find $\epsilon_{200}=4.6\times10^{-3}$. Having fixed all of the parameters in the fake lepton simulator using data from a single bin, we can extrapolate the results to all bins, and we show the result in Fig.~\ref{fig:trilepton_validation}. We see that the combination of prompt and fake lepton simulated backgrounds matches extremely well the backgrounds from Ref.~\cite{Chatrchyan:2014aea}. If we had not made the assumption that fake leptons originated exclusively from heavy-flavor jets, the relative rates of $\gamma^*/Z+$ jets and $t\bar t$ would change by an order of magnitude, and the agreement would no longer be acceptable.

Although not shown here, we also validated our fake lepton simulation against other ATLAS and CMS same-sign lepton searches \cite{ATLAS:2014kca,Khachatryan:2015gha}. In these analyses, only two final-state leptons were required (both of the same sign), and consequently contributions from multijet processes with \emph{two} fake leptons are also important. The fake simulation procedure described above also works well at reproducing the spectra from Refs.~\cite{ATLAS:2014kca,Khachatryan:2015gha} (except, on occasion, at the very high-$p_{\rm T}$ part of the background spectrum), but the overall normalization typically must be adjusted for each analysis. This is not surprising, since each analysis looks at a different final state (Ref.~\cite{ATLAS:2014kca} is an inclusive same-sign lepton search, while Ref.~\cite{Khachatryan:2015gha} requires two additional jets). It does suggest that the good shape agreement between MC simulation and CMS data shown in Fig.~\ref{fig:trilepton_validation} is robust provided we do not change the basic final-state topology considered in Ref.~\cite{Chatrchyan:2014aea}, and so we preserve the basic object selection from that analysis.